# FastImpute: A Baseline for Open-source, Reference-Free Genotype Imputation Methods - A Case Study in PRS313

Aaron Ge[1], Jeya Balasubramanian[1], Xueyao Wu[1], Peter Kraft[1], Jonas S. Almeida[1]

1 - Division of Cancer Epidemiology and Genetics, National Cancer Institute, National Institutes of Health, Maryland, USA

## Abstract

Genotype imputation enhances genetic data by predicting missing SNPs using reference haplotype information. Traditional methods leverage linkage disequilibrium (LD) to infer untyped SNP genotypes, relying on the similarity of LD structures between genotyped target sets and fully sequenced reference panels. Recently, reference-free deep learning-based methods have emerged, offering a promising alternative by predicting missing genotypes without external databases, thereby enhancing privacy and accessibility. However, these methods often produce models with tens of millions of parameters, leading to challenges such as the need for substantial computational resources to train and inefficiency for client-sided deployment.

Our study addresses these limitations by introducing a baseline for a novel genotype imputation pipeline that supports client-sided imputation models generalizable across any genotyping chip and genomic region. This approach enhances patient privacy by performing imputation directly on edge devices. As a case study, we focus on PRS313, a polygenic risk score comprising 313 SNPs used for breast cancer risk prediction. Utilizing consumer genetic panels such as 23andMe, our model democratizes access to personalized genetic insights by allowing 23andMe users to obtain their PRS313 score.

We demonstrate that simple linear regression can significantly improve the accuracy of PRS313 scores when calculated using SNPs imputed from consumer gene panels, such as 23andMe. Our linear regression model achieved an R² of 0.86, compared to 0.33 without imputation and 0.28 with simple imputation (substituting missing SNPs with the minor allele frequency). These findings suggest that popular SNP analysis libraries could benefit from integrating linear regression models for genotype imputation, providing a viable and light-weight alternative to reference based imputation.

**Availability**:

Web application: https://aaronge-2020.github.io/DeepImpute/

Code: https://github.com/aaronge-2020/DeepImpute

# 1. Introduction

Genotype imputation is essential in genomic research, as it enhances genetic data by predicting missing SNPs using reference haplotype information (Browning & Browning, 2007; Howie et al., 2009). Traditional methods leverage linkage disequilibrium (LD)—the tendency of nearby SNPs to be inherited together due to chromosome recombination. These methods infer untyped SNP genotypes by assuming similar LD structures between genotyped target sets and fully sequenced reference panels (Howie et al., 2009). However, to use traditional genotype imputation methods such as MiniMac4 (Das et al., 2016) or Beagle 5.4 (Browning et al., 2021), users must either rely on external services like the Michigan Imputation Server (Das et al., 2016), which can compromise data privacy, or download the entire reference genome onto a local machine, which may be time consuming and inefficient. Recently, reference-free deep learning-based methods (Mowlaei et al., 2023; Naito et al., 2021; Tanaka et al., 2022; Kojima et al., 2020) for genotype imputation have emerged as a promising alternative. They utilize machine learning to predict missing genotypes without external databases, enhancing privacy and accessibility by eliminating the need for data transfer to external servers.

Despite their promise, previous reference-free methods face significant challenges. They are not universally applicable across all genomic regions and often target the MHC region due to its difficulty to sequence given its high degree of polymorphism and structural variation (Naito et al., 2024). However, conducting genotype imputation in different regions requires retraining the entire model for that specific region. Given that neural network models often have tens of millions of parameters, substantial computational resources are required, making frequent retraining impractical. Additionally, these methods may lack the efficiency needed for deployment in client-side applications, undermining their privacy advantage.

Our primary contribution addresses these limitations by providing a baseline for zero-footprint, reference-free imputation methods. We introduce a novel genotype imputation pipeline that enables researchers to build client-sided imputation models, generalizable across any genotyping chip and, in theory, capable of predicting any genomic region. To establish this baseline, we trained a simple logistic regression model on phased data and a linear regression model on unphased data, though other models can also be utilized with the same procedures. As a case study, we focused on PRS313, a promising tool for breast cancer risk prediction leveraging genetic information across 313 SNPs (Mavaddat et al., 2019). By utilizing consumer-facing genetic panels such as 23andMe, our model democratizes access to personalized genetic insights, allowing individuals to obtain PRS313 scores more securely, accurately, and efficiently.

## 2. Methods

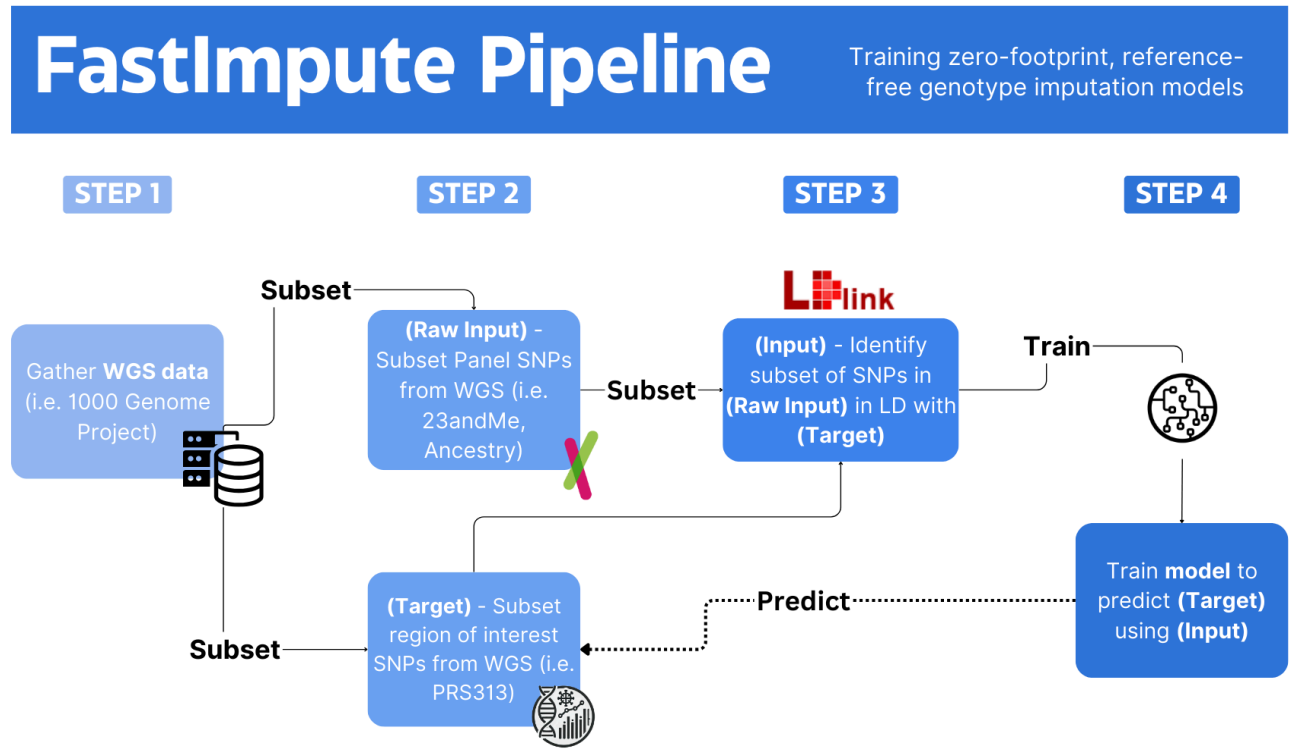

**Figure 1:** The FastImpute Pipeline.

The FastImpute pipeline is designed to train zero-footprint, reference-free genotype imputation models. The pipeline consists of four steps. **Step 1**: Gather Whole Genome Sequencing (WGS) data (e.g., from the 1000 Genomes Project). **Step 2**: Subset the panel SNPs from WGS data to include only those present in platforms like 23andMe and AncestryDNA, and a region of interest like PRS313. This subset serves as the raw input and target (output) for the model. **Step 3**: Using LDlink, identify a subset of SNPs in the raw input that are in linkage disequilibrium (LD) with the target SNPs (PRS313). These SNPs in LD with the target SNPs serve as the input for model training. **Step 4**: Train the model to predict the target SNPs using the input SNPs.

## 2.1 Determining the 23andMe SNPs on the V5 SNP

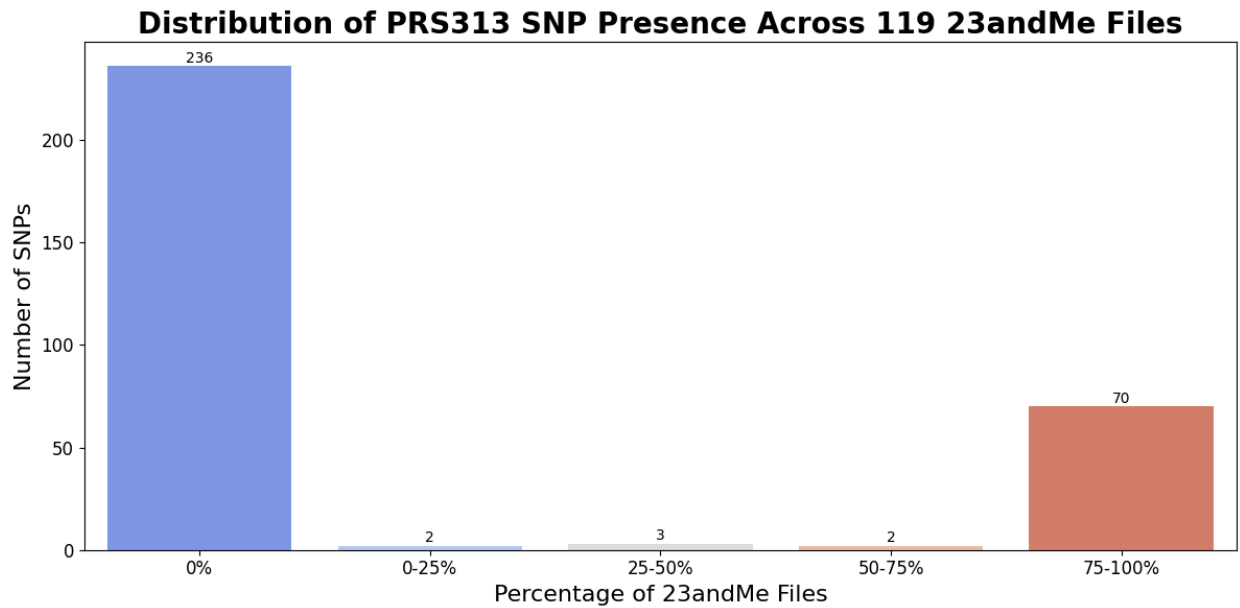

**Figure 2**: Distribution of PRS313 SNP Presence Across 119 23andMe Files

This bar chart shows the distribution of the presence of PRS313 SNPs across 119 23andMe V5 chip files. In total, 77 SNPs were found to be present within the 23andMe V5 chip. The x-axis represents the percentage of 23andMe files containing each SNP, divided into five ranges: 0%, 0-25%, 25-50%, 50-75%, and 75-100%. The y-axis indicates the number of SNPs within each range. A significant number of SNPs (236) are not present in any of the files (0%), while 70 SNPs are present in 75-100% of the files. The other ranges (0-25%, 25-50%, and 50-75%) contain very few SNPs, with counts of 2, 3, and 2, respectively. This distribution highlights the variability and inconsistency in SNP presence across different 23andMe V5 chip datasets.

      Since 23andMe does not publicly provide the positions of its V5 SNPs, we reverse-engineered the V5 chip using existing user data. Due to quality control measures, SNPs sampled from the same V5 chip can vary slightly, necessitating a method to ensure consistency across different datasets. Consequently, while 70 out of the total 77 PRS313 SNPs present in the 23andMe chip are found in most of the user data, there are 7 PRS313 SNPs that appear only sporadically (Figure 2).

To address this, we began by downloading all 23andMe data uploaded from 2023 onwards from the OpenSNP database (Greshake et al., 2014). We confirmed that the data generation dates were from 2022 onwards to ensure that none of the files were generated from an earlier version of the chip.

Next, we selected a reference 23andMe file known to be from the V5 chip to serve as a benchmark for filtering other 23andMe files. Since 23andMe uses a custom version of Illumina's Global Screening Array with 642,824 SNPs, we filtered the files to include only those containing between 600,000 and 700,000 positions. Additionally, we ensured that each sample shared at least 60% of its SNPs with the reference data. This step was crucial because some files labeled as 23andMe data might actually originate from other companies with different chipsets.

By applying these criteria, we filtered out datasets on OpenSNP that did not contain the V5 23andMe chip, leaving us with 119 23andMe files. The final step involved taking the union of the positions from all these files to accurately reconstruct the full 23andMe V5 chip panel.

## 2.2 Preparing the training Dataset

To build our genotype imputation model, we started by downloading the 1000 Genomes Project Data GrCh37 (Step 1, Figure 1) from the [official repository](#) (Auton et al., 2015). Next, using BCFtools (Li et al., 2011), we subsetted the target region of our genome imputation model, the PRS313 SNPs as reported by Mavaddat et al. (2019), and the SNPs we determined to be in the 23andMe panel from the 1000 Genomes Project data (Step 2, Figure 1). To reduce the likelihood of overfitting and improve model efficiency, we obtained linkage disequilibrium (LD) data for each PRS313 SNP (Step 3, Figure 1) using LD Proxy (Machiela et al., 2015). We focused on the subset of 23andMe SNPs that have an $R^2$ value greater than 0.01 with the PRS313 SNPs. For PRS313 SNPs that are multiallelic or not found in LD Proxy, we included all SNPs within a 500K base pair window in the 23andMe panel, resulting in 17,551 out of over 660,000 unique positions in the 23andMe V5 gene panel being used for training and evaluating the model.

We then converted the 17,551 positions in the 1000 Genomes data into allele dosages. Given the presence of multiallelic variants for some PRS313 SNPs, the 1000 Genome Project initially reports the specific variant inherited from the maternal and paternal sides (encoded as 0...n, for n number of alleles), rather than the allele dosage. To address this, we identified the reference and alternative alleles for each PRS313 SNP and converted all non-alternative alleles in the 1000 Genomes data to 0 and the alternative alleles to 1.

Finally, we created two versions of this data. The first version summed the allele dosages from the maternal and paternal sides to simulate unphased data. The second version kept the allele dosages from the maternal and paternal sides separate, maintaining the phased data format.

## 2.3 Model Training

This section explains step four of the FastImpute pipeline (Figure 1), in which we developed and evaluated two models: a logistic regression model to predict PRS313 SNPs from phased 23andMe genotype data and a linear regression model to predict PRS313 SNPs from unphased 23andMe genotype data. Both models underwent a consistent pipeline for hyperparameter tuning, training, and evaluation, ensuring comparability across different data types.

### 2.3.1 Data Loading and Preprocessing

Since the model is designed to capture linkage disequilibrium (LD) patterns, inter-chromosomal information is unnecessary for predicting SNP dosages. Therefore, we split the 23andMe panel data by chromosome, allowing us to construct separate models for each chromosome. Although the 23andMe panel already contains some of the PRS313 SNPs, each user's data varies slightly due to quality control measures, resulting in different numbers of PRS313 SNPs genotyped for each user. To address this, the output of our model is all PRS313 SNPs, including those already present in the user's 23andMe panel. This ensures that the model can output a predicted genotype for all 23andMe users, regardless of which PRS313 SNPs they are missing. Additionally, the inputs of our model are the SNPs of the 23andMe gene panel in LD with PRS313 SNPs.

## 2.3.2 Model Architecture and Training

The models trained in this pipeline are a linear regression model and a logistic regression model, both implemented using PyTorch (Paszke et al., 2019). The linear regression model consists of a single linear layer with L1 regularization (Tibshirani, 1996), which helps prevent overfitting by penalizing larger coefficients, acting as an implicit filter for SNPs with low linkage disequilibrium (LD) to PRS313 SNPs. The logistic regression model includes an additional sigmoid layer appended to the linear layer.

We trained the models using a randomly shuffled 80/20 data split, where 80% (2003 samples) of the data is used for training and validation, and the remaining 20% (501 samples) is reserved for testing. We employed the Optuna library (Akiba et al., 2019) for hyperparameter tuning, utilizing 10-fold cross-validation to optimize the model's parameters across 50 trials for each chromosome.

## 2.3.3 Evaluation Metrics

To evaluate the imputation performance for each of our 22 genotype imputation models (one for each chromosome), we computed several metrics: $R^2$, Imputation Quality Score (IQS) (Lin et al., 2010), accuracy, and AUC. $R^2$ was calculated using the Scikit-learn library (Pedregosa et al., 2018) to indicate the proportion of variance in the true genotypes explained by the imputed dosages. AUC (Area Under the ROC Curve) was computed using PyTorch (Paszke et al., 2019) to assess the discriminative ability of the imputed dosages.

Since imputing allele dosages is a multi-class classification problem (with allele dosages of 0, 1, or 2), we calculated the one-vs-all AUC for each class for the linear regression model trained with unphased data. We then computed the mean AUC of the three classes. In cases where there were no positive classes for a genotype, resulting in an undefined AUC value, these undefined values were excluded when calculating the mean AUC.

Each metric was computed on the test set of each chromosome, and the median values across all 22 chromosomal models are reported in Figure N. Additionally, the $R^2$ of each individual SNP within PRS313 was computed for both the linear regression and logistic regression models. The distributions of these $R^2$ values are plotted in Figure N and compared with the $R^2$ values of Beagle (Browning et al., 2021).

## 2.4 Benchmarking Beagle

To benchmark the performance of Beagle 5.4 (Browning et al., 2021), we left out the same 501 samples that we used to evaluate our previously trained models from the 1000 Genome Project to serve as the test set. We then ran Beagle 5.4 on the full 23andMe panel data, excluding the overlapping PRS313 SNPs that are already present in the panel, using the remaining 2003 samples from the 1000 Genomes Project as the reference genome. The imputation results were subsequently evaluated for $R^2$, IQS, Accuracy, and AUC ROC.

## 2.4 Deployment

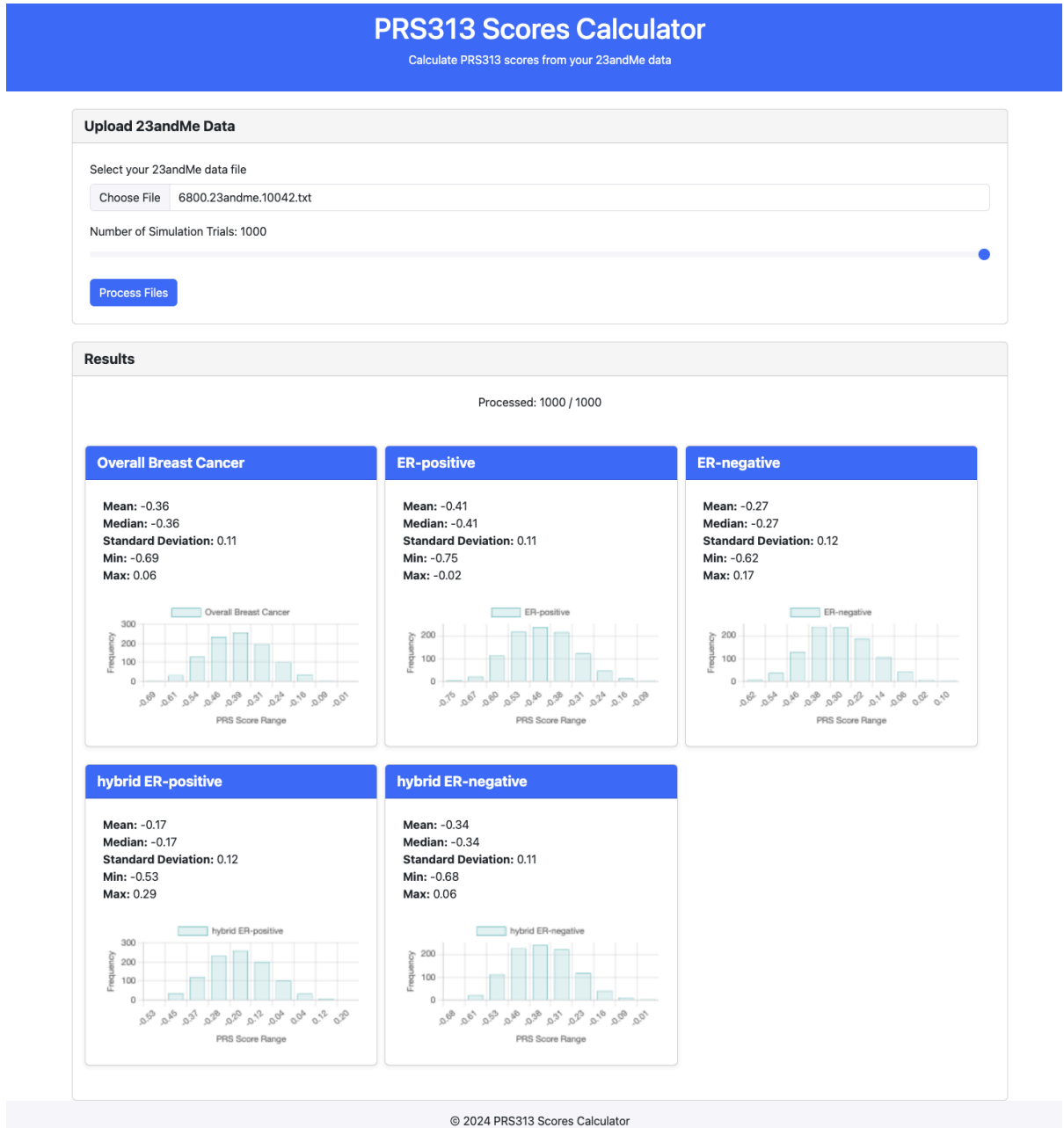

**Figure 3**: The PRS313 Scores Calculator interface.

This website displays results of the polygenic risk score (PRS) calculations for various breast cancer phenotypes based on user 23andMe genotype data. Users can upload their 23andMe data file and specify the number of simulation trials to process their PRS313 scores. The results section displays the processing status and the calculated PRS scores for five breast cancer phenotypes: Overall Breast Cancer,

ER-positive, ER-negative, hybrid ER-positive, and hybrid ER-negative. Each phenotype panel includes statistical summaries of the PRS scores, such as mean, median, standard deviation, minimum, and maximum values, along with a histogram showing the distribution of PRS scores across the simulation trials. For more information, please see section 2.4.

We deployed our linear regression (unphased) model on GitHub at https://aaronge-2020.github.io/DeepImpute/, allowing users to privately and securely calculate their PRS313 scores on any devices, including their smartphones. The reason we deployed this model over the logistic regression model despite its better performance is because 23andMe data does not contain phasing information. To calculate the polygenic risk score (PRS) from a user's 23andMe genotype data, we conducted the following steps:

First, the user's 23andMe data was converted from genotypes to allele dosages by referencing the 1000 Genomes Project for the reference and alternate alleles at each position. Due to quality control measures by 23andMe, different users have data with slightly different positions, leading to varying missing data across users. To address this, we simulated missing SNP values based on minor allele frequency (MAF) data taken from the 1000 Genome project (1000 Genome Consortium, 2015). This approach allowed us to calculate PRS313 risk scores across multiple simulations, generating a distribution of potential risk scores based on the missing data from each user's 23andMe panel. Users could specify the number of simulations, balancing the trade-off between accuracy and computational time. Typically, running 1,000 simulations only takes approximately 7-10 seconds based on the speed of the user's machine, due to the model's small size and portability.

For each simulation, two random draws were conducted for each allele, with the probability of the allele being present based on the MAF, to determine the dosage for each allele. These dosages were then summed to simulate unphased data. The simulated data from each simulation was passed into the imputation model to impute the PRS313 SNPs.

Since the user's 23andMe data may already contain certain PRS313 positions, the allele dosages at already genotyped locations were used in place of the imputed dosages. The PRS313 SNPs dosages from each simulation were then used to calculate PRS scores. Beta values for each

SNP, corresponding to different breast cancer phenotypes, were retrieved from an external dataset. The PRS scores were computed by multiplying the imputed dosages by their respective beta values and summing these products for each phenotype. This process was carried out for each SNP and aggregated to generate a cumulative PRS for each phenotype.

The results from these simulations were aggregated, and statistical summaries, including mean, median, standard deviation, minimum, and maximum values, were computed for each phenotype. These results were visualized using binned histograms to display the distribution of PRS scores, providing a variance estimate for the PRS score.

# 3. Results

The performance of different genotype imputation methods was evaluated using $R^2$, IQS, AUC, and accuracy to determine their effectiveness in genotype imputation and predicting polygenic risk scores (PRS). Our analysis, presented in Figures 4 through 7, has shown that although Beagle (Browning et al., 2021) consistently performed the strongest across various metrics and PRS phenotypes, the baseline linear models do not fall significantly behind. This is surprising, given the simplicity of the models and the linear regression model's lack of use of a phased reference genome.

## 3.1 Evaluation of Genotype Imputation Models

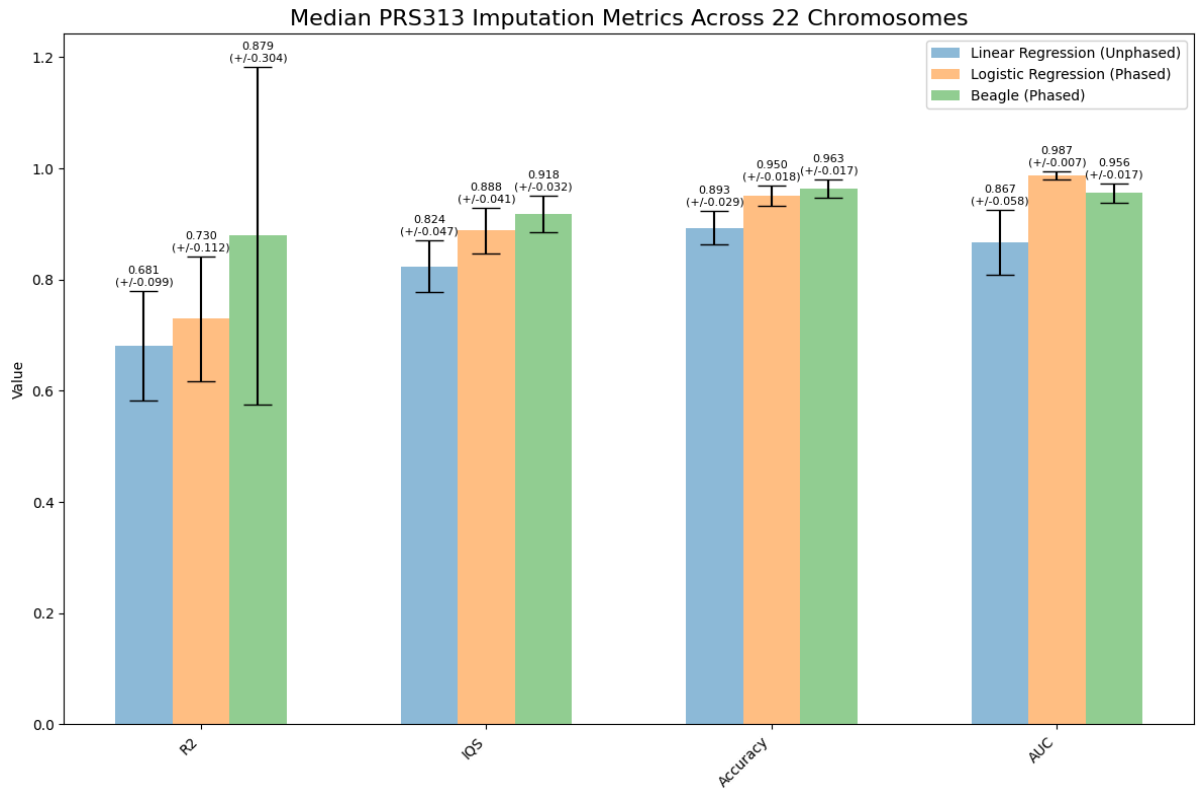

**Figure 4:** Median PRS313 Imputation Metrics Across 22 Chromosomes for Different Methods.

This bar plots display the median evaluation metrics (R², IQS, Accuracy, and AUC) across 22 chromosomal models for three genotype imputation methods: Linear Regression trained on unphased data, Logistic Regression trained on phased data, and Beagle using phased data. The metrics for each chromosome model were first calculated, and then the median values across all 22 chromosomes were determined. Error bars represent the standard deviation of the metrics across the different chromosomes.

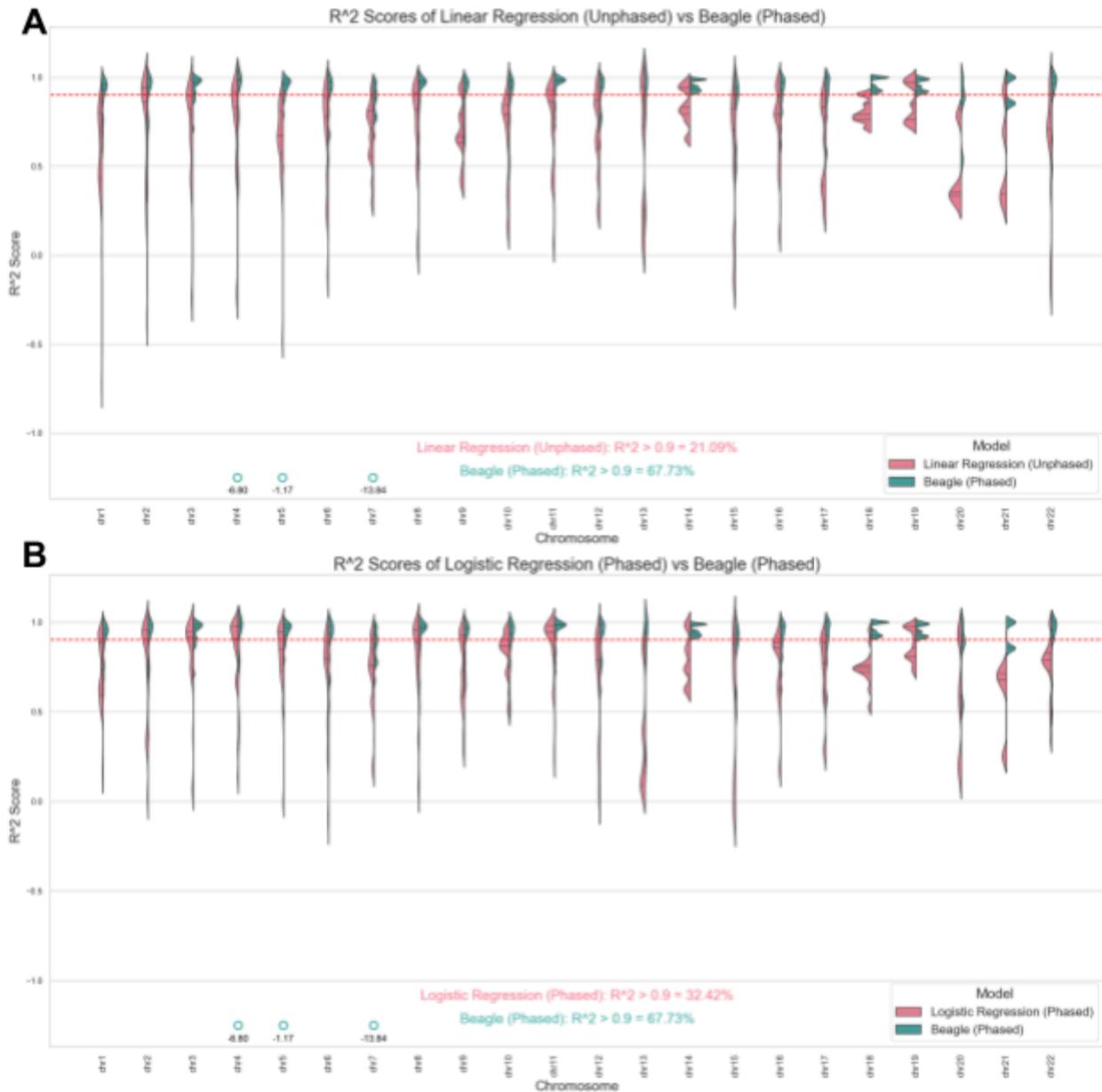

**Figure 5:** R² Score Distributions for SNP Genotype Imputation Models across Chromosomes.

**5A** shows the R² scores of SNPs imputed using the Linear Regression model trained on unphased data vs. Beagle results imputed using phased data. The violin plots illustrate the distribution across 22 chromosomes. The Beagle model shows a higher proportion of SNPs with R² > 0.9 (68.39%) compared to the Linear Regression model (21.71%). **5B** shows the R² scores of SNPs imputed using the Logistic Regression trained on phased data vs. Beagle results imputed using phased data. Similar to (A), the

Beagle model outperforms the linear method, with 68.39% of SNPs exceeding R² > 0.9, compared to 32.90% for Logistic Regression. Outliers are plotted individually and labeled.

As shown in Figure 4, which compares the median metrics (R², Imputation Quality Score (IQS), Accuracy, and AUC) of each model across 22 chromosomes, Beagle (Browning et al., 2021) consistently achieves the highest median values. When assessed using IQS, accuracy, and AUC, Logistic Regression (phased) and Linear Regression (unphased) performs very comparably to Beagle. While the Logistic Regression (phased) achieved a median IQS of 0.888 +/- 0.041 and the Linear Regression (unphased) had an IQS of 0.824 +/- 0.047, Beagle has a median IQS of 0.918 +/- 0.032. However, when assessed using $R^2$, beagle performs significantly stronger than the linear methods, with an $R^2$ of 0.879 +/- 0.304, compared to 0.730 +/- 0.112 of logistic regression and 0.681 +/- 0.099 of linear regression. The high standard deviation in the $R^2$ of Beagle is due to the presence of two problematic SNPs, which achieved extremely negative $R^2$ values (chr7:55192256 and chr4:84370124). The significance of phasing information in genotype imputation is also highlighted in Figure 4, which illustrates the stronger performance of the logistic regression model using phased data compared to the linear regression model using unphased data.

The distribution of R² scores across chromosomes, shown in Figure 5, further illustrates the performance differences between imputation methods. Beagle exhibits a high proportion of SNPs with R² > 0.9 (68.39%), significantly outpacing Linear Regression (unphased), which only achieves this threshold for 21.71% of SNPs. Similarly, Logistic Regression (phased) reaches this level for 32.90% of SNPs, highlighting its stronger performance over the unphased model. These findings underscore that there is indeed a significant advantage to using phased data and reference-base methods like Beagle. However, they also highlight the decent efficacy of the baseline models, especially given their simplicity and reference-free nature.

## 3.2 Evaluation of PRS Scores Accuracy with Imputed Genotypes

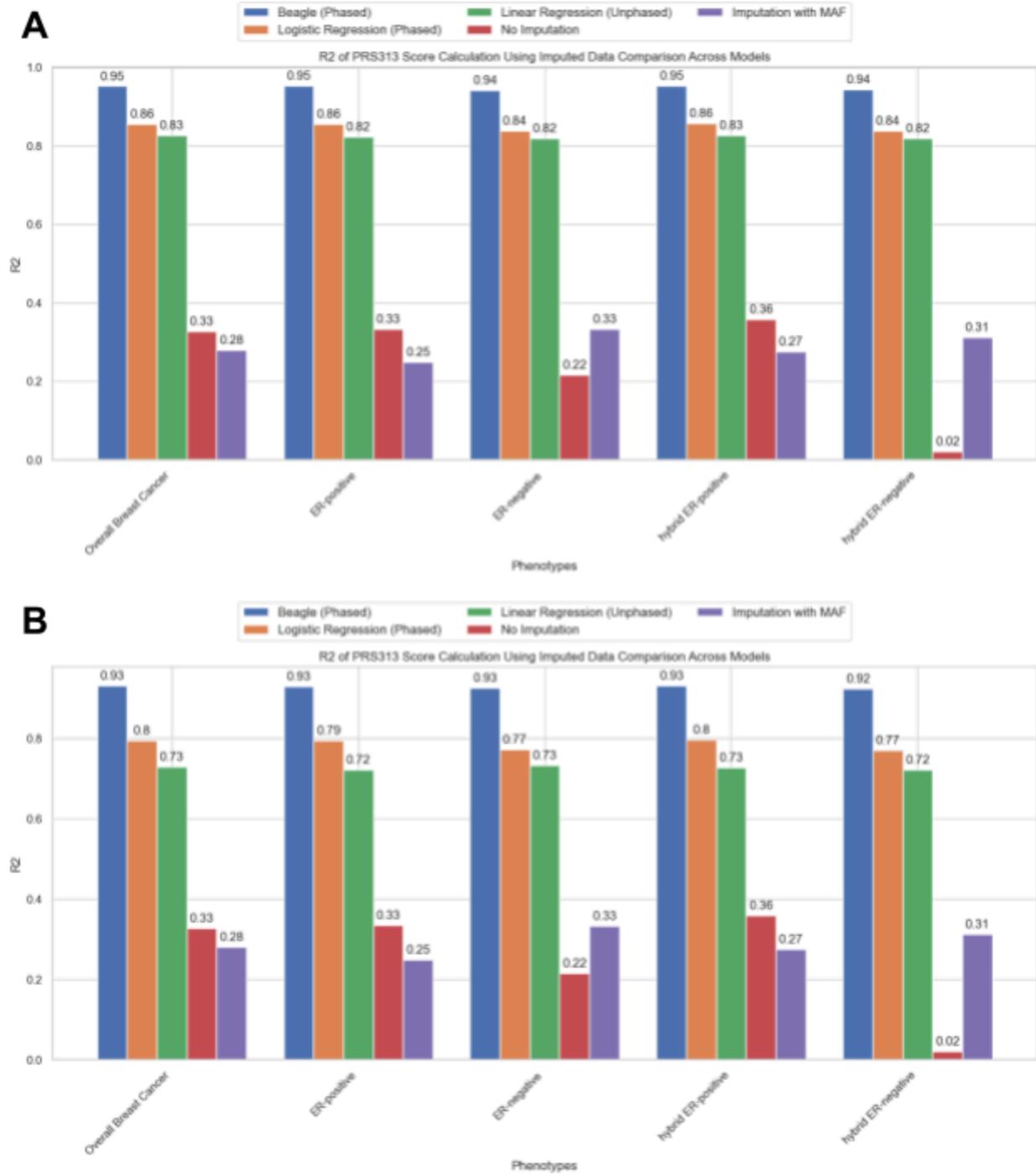

**Figure 6:** Comparison of the $R^2$ of the PRS score Across Different Models.

The $R^2$ values were calculated by comparing the PRS scores calculated using imputed genotypes versus the PRS score calculated using the real genotypes obtained from the WGS data. Due to quality control measures, although there are 77 SNPs present within the 23andMe genotyping panel, user data may have varying numbers of PRS313 SNPs. **6B** shows the R² values, when the PRS313 scores is calculated only using imputed PRS313 dosages while **6A** shows R² values when all 77 SNP positions are known.

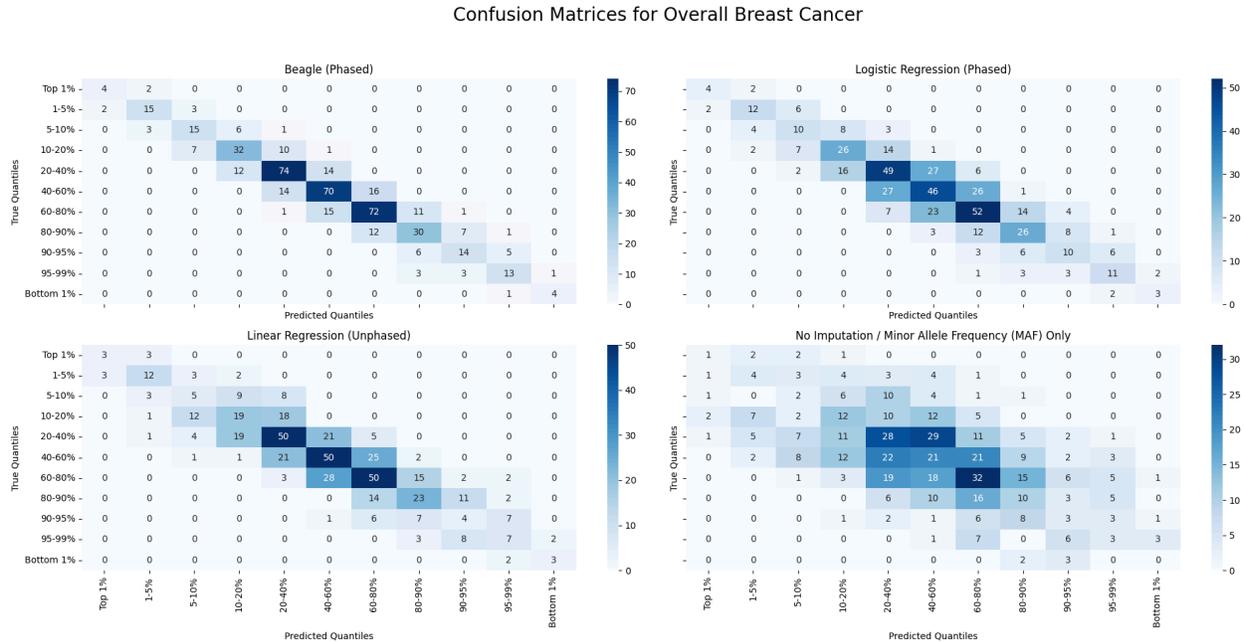

**Figure 7:** Confusion Matrices for Overall Breast Cancer Risk Prediction.

The figure shows confusion matrices comparing the performance of different imputation methods for predicting overall breast cancer risk using PRS313 scores. The matrices display the agreement between true quantiles and predicted quantiles across various imputation techniques. Each confusion matrix's x-axis represents the predicted quantiles, while the y-axis represents the true quantiles. Darker shades indicate a higher number of samples falling into the respective quantile categories, highlighting the distribution of prediction accuracy across the different methods.

The R² values, depicted in Figure 6, compare PRS scores calculated using imputed genotypes to those calculated with real genotypes. Beagle (Browning et al., 2021) achieves the highest R² values across all phenotypes, with an R² of 0.95 for overall breast cancer, indicating a strong correlation between imputed and actual genotypes. However, it's noteworthy that the baseline models—Logistic Regression (phased) and Linear Regression (unphased)—also perform surprisingly well. For instance, Logistic Regression achieves an R² of 0.86 for overall breast cancer, only slightly lower than Beagle. These models significantly outperform the PRS score calculated without imputation, with an $R^2$ of 0.33 and the Imputation with MAF method, which shows even lower R² values, with an R² of only 0.28.

The confusion matrices in Figure 7 illustrate the agreement between true and predicted quantiles for overall breast cancer risk using PRS313 scores. It is important to note that when imputing with MAF values, the same set of dosages are imputed for each sample, regardless of an individual's genotype. Thus, the relative quantile of an individual within the test set will not change, making the confusion matrix for PRS313 score quantiles identical to using no imputation when imputing with only the MAF.

As shown in Figure 7, out of the six subjects in the top 1% of PRS scores, both the logistic regression model and Beagle correctly classified 4 out of 6. The linear regression model correctly classified 3 out of 6, while imputing with MAF only led to 1 out of 6 being correctly classified. For the misclassified patients in the top 1% of PRS scores, Beagle, linear regression, and logistic regression placed all of them in the next highest score quantile (1-5%). In contrast, imputing with MAF placed 2 out of 6 patients in the 5-10% quantile and 1 out of 6 in the 10-20% quantile.

The discrepancy in accuracy becomes even more apparent in the 1-5% quantile, where imputing with MAF correctly classified only 4 out of 20 patients, with almost half (8 out of 20) falling outside the top 20% quantile. Beagle correctly classified 15 out of 20, and both linear and logistic regression models correctly classified 12 out of 20, with misclassified samples only found within adjacent quantiles.

## 4. Discussion

This study introduces lightweight, client-sided genotype imputation models that enhance both the accuracy and accessibility of polygenic risk score calculations. Our approach addresses key challenges in existing imputation methods, balancing performance, computational efficiency, and privacy. The ability to run these models on edge devices in real-time represents a significant advancement in point-of-care genetic testing and personalized medicine.

Traditional reference-based methods, while accurate, are computationally intensive and limited by reference panel accessibility. The size of the GRCh37 1000 Genome Project files (14.8 GB) can be prohibitive for many researchers, necessitating reliance on services like the Michigan Impute Server (Das et al., 2016). These services lack real-time genotype imputation capabilities and can cause delays in research and clinical decision-making due to server downtimes and long queue times. Additionally, these methods may underperform when target sets differ from discovery panels (Levi et al., 2024), as shown by the variability in Beagle's $R^2$ in our results, particularly at a few specific PRS313 SNPs (chr7:55192256 and chr4:84370124). This variability could lead to inconsistent genetic risk assessments in underrepresented patient populations. Reference-free approaches have been developed to address these issues (Duong et al., 2023; Mowlaei et al., 2023; Naito et al., 2021; Tanaka et al., 2022; Kojima et al., 2020). However, they are often too large for browser-based imputation and require computationally expensive retraining for imputing on different regions, limiting their practical clinical application.

Our method offers an efficient, accurate, and privacy-preserving alternative that, to our knowledge, is the first serverless genotype imputation model. Using a baseline linear regression model, our solution is simple yet effective, with significantly less expensive retraining compared to deep neural networks. This adaptability suits various genomic regions and pipelines while enabling data processing on edge devices like smartphones.

As demonstrated by our PRS313 case study for breast cancer risk assessment, the clinical implications are significant: using our pipeline, healthcare providers could, in theory, perform real-time genetic risk assessments during patient consultations, seamlessly integrating genetic information into clinical decision-making. By preserving user privacy and enhancing accessibility, our approach could democratize access to gene imputation, enabling its integration into routine clinical practice even in resource-limited settings. This could lead to earlier disease detection, more personalized treatment plans, and improved patient outcomes across a wide range of genetic conditions. According to Mavaddat et al. (2019), women in the top 1% of the PRS313 distribution have a predicted risk approximately four times larger than those in the middle quintile, aligning with the UK NICE definition of high risk for breast cancer. Our logistic regression model using phased data performs comparably to Beagle in identifying high-risk individuals. In addition, the linear regression model, which does not have access to phasing information, only misclassified one additional patient. Given its elimination of server-related privacy issues and theoretically zero-cost scalability, it shows promise for widespread, routine PRS calculations.

Despite its potential, our approach has some limitations. First, our method of reverse engineering the 23andMe panel assumes that the 119 users we sampled provide a comprehensive overview of the V5 chip. Although this sample size is relatively small, it represents almost half of all publicly available 23andMe data on openSNP.org from 2020 onwards, the release date of the V5 23andMe chip. Given our limited resources and the scarcity of publicly available data, this selection minimizes the potential impact on the validity of our results. Future research may collaborate with consumer genetic testing companies to obtain more extensive and representative data on genotyping chips, ensuring a more robust analysis.

Second, our data selection method focuses on regions within a +/- 500K base pair window with linkage disequilibrium (LD), following the default values provided by LDProxy. This approach captures significant regions but might overlook other regions with high LD and informative locations, potentially missing some ancestry information contained in the original 23andMe panel that could affect the accuracy of polygenic risk scores (PRS) due to varying LD structures across ancestries. While we limited feature selection to these regions to prioritize small model size and portability, future studies should consider expanding the window size during data selection to capture additional genetic variation and evaluate imputation effectiveness across different genomic regions and polygenic risk scores.

Third, the PRS calculated in our web application are not calibrated based on the user's superpopulation. This calibration is important because the genetic distance between the user and our training dataset can influence the predictive power of our genotype imputation models. However, due to the limited availability of diverse training datasets, we were unable to implement superpopulation-specific calibration. Future research should focus on training separate imputation models for each superpopulation to enhance imputation accuracy and normalize the PRS scores accordingly. Additionally, researchers can consider exploring more complex models, such as neural networks, to reach a balance between performance and portability.

## 5. Conclusion

Our study demonstrates the viability of lightweight, client-sided genotype imputation models, highlighting their potential for preserving user privacy and enhancing accessibility. These results have practical implications, particularly for breast cancer risk prediction. Our baseline models identified high-risk individuals using 23andMe data with comparable performance to Beagle, underscoring their potential utility in clinical settings.

## Funding

This work was funded by the National Cancer Institute (NCI) Intramural Research Program (DCEG/Episphere #10901).